\documentclass[12pt]{article}
\usepackage[a4paper, margin=1in]{geometry}
\usepackage{authblk}
\usepackage{hyperref}
\usepackage{amsmath}
\usepackage{graphicx}
\usepackage{listings}
\usepackage{tikz}
\usetikzlibrary{arrows.meta, positioning, shapes.multipart}
\usetikzlibrary{calc}
\usepackage{graphicx}
\usepackage{booktabs}      % tables
\usepackage{siunitx}       % for aligning numbers
\usepackage{pgfplots}      % for the bar chart
\pgfplotsset{compat=1.17}  % or your installed version

\title{Evaluating Large Language Models for Workload Mapping and Scheduling in Heterogeneous HPC Systems}
\author[1]{Aasish Kumar Sharma}
\author[1]{Julian Kunkel}
\affil[1]{Faculty of Mathematics and Computer Science, Georg-August-Universität Göttingen, Germany}

\date{July 2025}

\usepackage[table,xcdraw]{xcolor}

\begin{document}
\maketitle

\begin{abstract}
\noindent
Large Language Models (LLMs) are increasingly investigated for technical reasoning, yet their capacity to perform structured, constraint-based optimization purely from natural-language input remains insufficiently understood. 
This study evaluates 21 publicly available LLMs on a representative heterogeneous High-Performance Computing (HPC) workload mapping and scheduling problem. 
Each model receives the same textual description of system nodes, task requirements, and scheduling constraints, and must assign tasks to nodes, compute the total makespan, and explain its reasoning. 
A manually derived analytical optimum of 9\,h\,20\,s serves as the ground-truth reference.

Three models (\texttt{o3-mini}, \texttt{GPT-4.1 Mini}, and \texttt{Gemini Pro 2.5}) exactly reproduced the analytical optimum while satisfying all constraints, whereas twelve additional models achieved near-optimal results within two minutes of the reference. 
The remaining models produced suboptimal schedules, often due to arithmetic inaccuracies or dependency violations. 
All models achieved full task coverage and feasible node mappings, though only about half maintained strict constraint adherence. 
19 generated partially executable verification code, and 18 provided coherent step-by-step reasoning, demonstrating strong interpretability even when logical errors occurred.

These results delineate the current capability boundary of LLM reasoning in combinatorial optimization: leading models can reconstruct optimal schedules directly from natural-language descriptions, but most still struggle with precise timing, data-transfer arithmetic, and dependency enforcement. 
The findings underscore LLMs’ promise as explainable co-pilots for optimization and decision-support tasks, rather than autonomous solvers.

\end{abstract}

\begin{keywords} 
Large Language Models (LLMs), HPC Scheduling, Constraint Reasoning, Makespan Optimization, Explainable AI 
\end{keywords}

\section{Introduction}

Heterogeneous High-Performance Computing (HPC) systems that integrate CPUs, GPUs, and domain-specific accelerators form the backbone of modern scientific computing and large-scale data processing~\cite{deelman2025high,brodtkorb2010state}. Efficient workload mapping and scheduling across such diverse resources are essential to minimize makespan, maximize throughput, and ensure balanced utilization. However, the scheduling problem is inherently combinatorial and NP-hard, typically addressed through mathematical programming, heuristics, or metaheuristics~\cite{topcuoglu2002performance,kwok1999static,sharma2025review}. While these classical methods can yield optimal or near-optimal solutions for small-scale instances, they require explicit system modeling, precise mathematical formulations, and domain expertise, making them less accessible and computationally intensive for dynamic real-world scenarios.

Recent advances in Large Language Models (LLMs), such as GPT-4, Claude, and Qwen, have shown that they can perform symbolic reasoning, follow complex instructions, and generate structured code directly from natural-language input. These emerging capabilities raise a compelling research question:

\begin{quote}
\emph{Can LLMs reason about multi-constraint optimization problems, such as HPC workload mapping and scheduling, purely from natural-language descriptions, without relying on formal equations or optimization solvers?}
\end{quote}

To address this question, we designed a capability evaluation experiment that isolates reasoning quality rather than algorithmic performance. Each LLM is provided with a full natural-language description of a heterogeneous HPC system, including nodes, resources, task requirements, dependencies, and objectives. Without any structured or mathematical input, the model must produce a valid mapping and scheduling plan, compute the overall makespan, and explain its reasoning process. The generated solutions are compared against a \textit{manually derived and analytically verified optimal schedule} that serves as the ground truth. This setup enables a direct assessment of whether an LLM can reconstruct the logic of constrained optimization and emulate human-style reasoning in HPC scheduling contexts.

\vspace{0.5em}
\noindent \textbf{Research Questions}
\begin{itemize}
    \item \textbf{RQ1:} Can LLMs comprehend and reason over natural-language descriptions of resource, feature, and dependency constraints in HPC scheduling problems?
    \item \textbf{RQ2:} Can they independently derive optimal or near-optimal schedules and compute accurate makespans?
    \item \textbf{RQ3:} What reasoning errors and failure patterns emerge, and what do they reveal about the current limits of LLMs’ structured optimization reasoning?
\end{itemize}

\vspace{0.5em}
\noindent \textbf{Contributions.}
This paper bridges HPC workload mapping and scheduling with AI reasoning by:
(i) introducing a manually validated framework to assess the capability of Large Language Models (LLMs) to perform constraint reasoning from natural-language descriptions of systems and workloads on a single prompt; 
(ii) conducting a systematic evaluation of 21 publicly available models using a common analytical benchmark; 
(iii) establishing a taxonomy of typical reasoning and constraint-violation patterns that characterize LLM-based optimization attempts; and 
(iv) defining the current capability boundary of LLMs in structured reasoning and positioning them as explainable co-pilots for future hybrid human–solver scheduling systems.

\vspace{0.5em}
The remainder of this paper is organized as follows. 
Section~\ref{sec:litreview} reviews related work on traditional and AI-assisted scheduling approaches. 
Section~\ref{sec:methodology} details the proposed evaluation framework, benchmark setup, and prompt design. 
Section~\ref{sec:findinganddiscussion} presents the empirical findings and comparative analysis across models along with discusses observed reasoning behaviors and failure patterns. 
Finally, Section~\ref{sec:conclusionandfuturework} concludes the study and outlines directions for advancing LLM-assisted optimization in heterogeneous HPC systems.

\section{Literature Review}
\label{sec:litreview}

\subsection{Classical Optimization Approaches}
    
Workload mapping and scheduling in HPC has a long history of research, with prominent approaches including integer and mixed-integer linear programming (ILP/MILP), heuristics such as Heterogeneous Earliest Finish Time (HEFT) and Opportunistic Load Balancing (OLB), and metaheuristics (e.g., Genetic Algorithms (GA), Simulated Annealing (SA)) \cite{topcuoglu2002performance, kwok1999static, hussain2013survey}. These methods are valued for their guarantees of feasibility and, in the case of mathematical programming, provable optimality for small-to-medium-scale instances \cite{sharma2025workflow}.

%\jk{ADD here som of the graphs you produced before. citing that paper.}
\begin{figure}[ht!]
    \centering
    \includegraphics[width=0.92\linewidth]{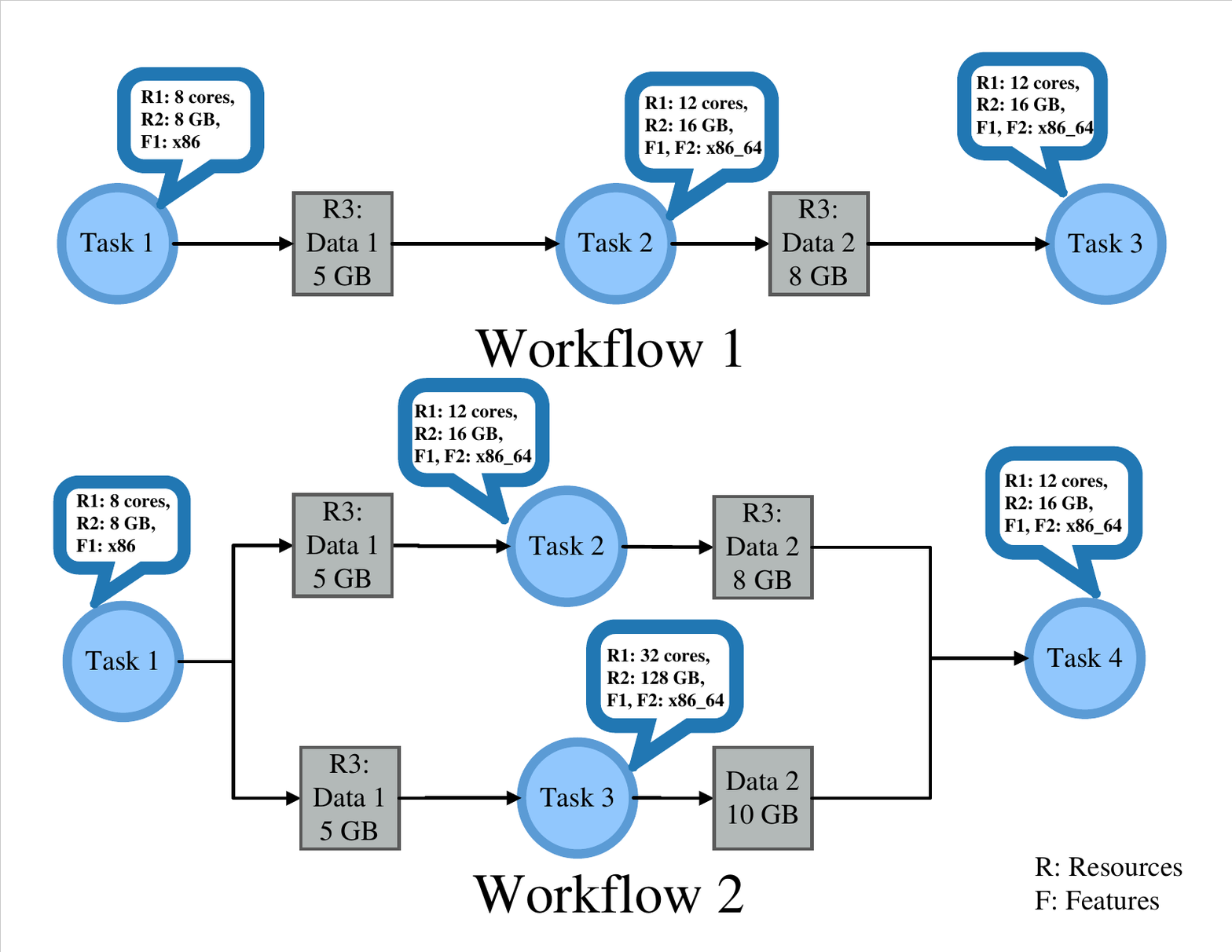}
    \caption{Simplified Petri Net representing task dependencies and resource tokens \cite{sharma2025workflow}.}
    \label{fig:PetriNetWithResourceAndFeatureRequest}
\end{figure}

Recent work has introduced Graph Neural Networks (GNNs) and Reinforcement Learning (RL) for dynamic, adaptive scheduling in heterogeneous and time-varying contexts \cite{ye2024deep, grinsztajn2021readys, sharma2025grapheonrl}. Figure~\ref{fig:PetriNetWithResourceAndFeatureRequest} represents a standard Petri net workflow with process and state representation.

\subsection{LLMs for Optimization and Scheduling}

There is a growing body of work exploring LLMs for optimization tasks. Wang et al. \cite{cui2025large} propose a benchmark suite evaluating LLMs' performance in code optimization and scheduling. Gupta et al. \cite{nichols2024hpc} introduce HPC-Coder, a model for automated code annotation and parallelism suggestion. LLM-based agents are increasingly proposed as planners or co-pilots for scheduling, resource allocation, and configuration tasks.

However, direct application of LLMs to multi-constraint, large-scale HPC scheduling remains under-explored, with known challenges in context length, mathematical precision, and constraint adherence. While classical scheduling benchmarks exist for algorithmic and solver-based approaches, there is currently no standardized methodology for evaluating LLMs’ reasoning ability in HPC workload mapping from natural-language descriptions, motivating the present study.

To date, no published work has directly evaluated LLMs’ intrinsic reasoning ability to perform scheduling and makespan computation from natural-language instructions alone. The present study fills this gap by providing a manually validated benchmark of reasoning correctness and constraint adherence.

\section{Methodology}
\label{sec:methodology} 

\subsection{Benchmarking Framework}

\begin{figure}[h!]
\centering
\begin{tikzpicture}[
    node distance=1.2cm and 1.7cm,
    box/.style = {rectangle, draw, rounded corners, minimum width=2.3cm, minimum height=0.8cm, align=center, fill=gray!10},
    proc/.style = {rectangle, draw, thick, fill=blue!10, minimum width=2.4cm, minimum height=0.9cm, align=center},
    output/.style = {rectangle, draw, fill=orange!10, minimum width=2.3cm, minimum height=0.8cm, align=center},
    compare/.style = {rectangle, draw, thick, dashed, minimum width=2.7cm, minimum height=0.9cm, align=center, fill=yellow!25},
    arrow/.style = {thick, -{Stealth[scale=1.1]}, shorten >=2pt, shorten <=2pt}
  ]
% Input box
\node[box] (input) {System \& Workload\\ Scenario};

% Methods
\node[proc, above left=of input, yshift=0.0cm, xshift=-0.2cm] (milp) {Classical Solver};
\node[proc, above right=of input, yshift=0.0cm, xshift=0.2cm] (llm) {LLM (GWDG Hosted)};

% Outputs
\node[output, above=of milp, yshift=-0.3cm] (milpout) {Schedule \&\\ Reasoning};
\node[output, above=of llm, yshift=-0.3cm] (llmout) {Schedule \&\\ Reasoning};

% Comparator
\node[compare, above=1.7cm of $(milpout)!0.5!(llmout)$] (compare) {Evaluation \&\\ Comparison};

% Arrows from input to methods
\draw[arrow] (input.north west) -- (milp.south);
\draw[arrow] (input.north east) -- (llm.south);

% Arrows from methods to outputs
\draw[arrow] (milp.north) -- (milpout.south);
\draw[arrow] (llm.north) -- (llmout.south);

% Arrows from outputs to compare
\draw[arrow] (milpout.north) -- (compare.south west);
\draw[arrow] (llmout.north) -- (compare.south east);

\end{tikzpicture}
\caption{Evaluation methodology: Each LLM receives the same natural-language description of the HPC system and workload as used in the manually derived optimal baseline. Outputs are compared for constraint adherence, correctness of makespan, and quality of reasoning and explanation.}

\label{fig:framework}
\end{figure}
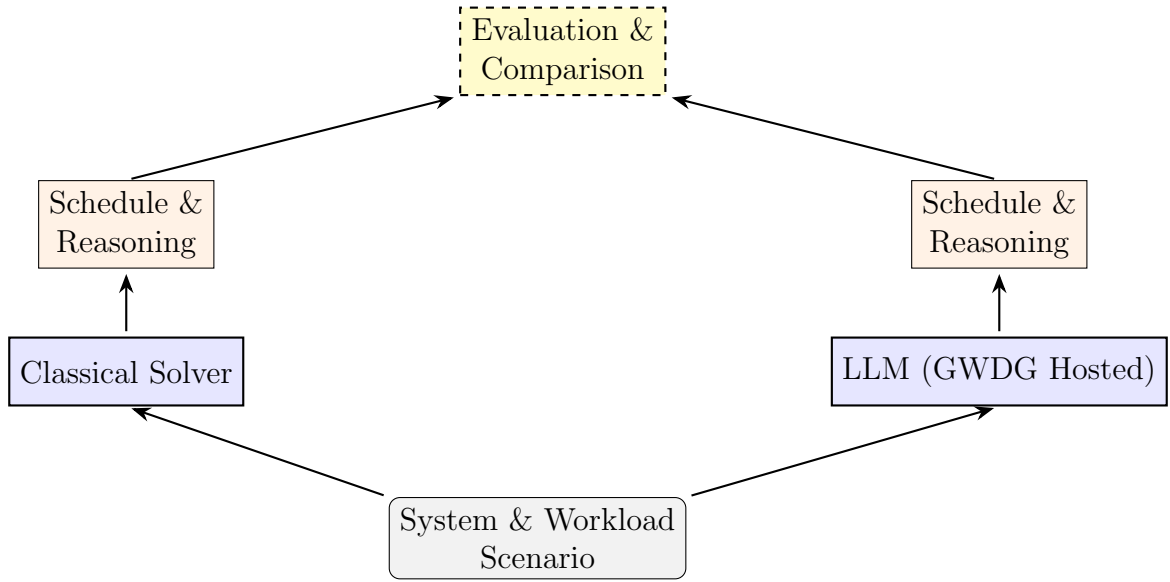

We design an evaluation framework (Fig.~\ref{fig:framework}) to test whether LLMs can internally reason about mapping and scheduling constraints and objectives solely from natural-language instructions. Unlike solver-based benchmarks, the reference task-to-node map and schedule in our setup is \textit{manually derived} through analytical reasoning and validated step-by-step for feasibility, optimality, and makespan. LLMs are provided with the same textual description of the scenario-no equations, JSON input, or symbolic hints, and must independently reconstruct and solve the problem. Their responses are evaluated for constraint satisfaction, correctness of makespan computation, and clarity of explanation.

The modeling details as well as the mathematical and the heuristics solvers are based on \cite{sharma2025workflow}. Besides, LLMs are prompted with the full scenario in natural language, and asked to generate task-to-node assignments, start/end times, and reasoning. Solutions are compared to those from a MILP implementation (Python/PuLP) and a heuristic baseline (e.g., HEFT). \cite{jadhav2025evaluating} presents, LLMs provides better solution when refined prompts are looped with proper reasoning which could be another analysis for us, while for current benchmarking we do not loop.

Table~\ref{tab:assessment} summarizes our qualitative assessment of different LLMs\footnotetext[1]{See \cite{gwdgchataimodels} for the current list of models hosted at GWDG.} in scheduling HPC workloads. Each LLM model was prompted with the same system and workload scenario. The criteria include overall makespan, task and system specification correctness, quality of reasoning, explanation clarity, code generation, and mapping validity.

\subsection{Sample Scenario}

A representative test scenario:

\begin{itemize}
    \item \textit{Nodes}:
    \begin{itemize}
        \item NodeA: 32 CPUs, 128 GB RAM, Features: [CPU, GPU], Data Rate: 10 Gbps
        \item NodeB: 64 CPUs, 256 GB RAM, Features: [CPU], Data Rate: 5 Gbps
        \item NodeC: 16 CPUs, 64 GB RAM, Features: [CPU, SSD], Data Rate: 2 Gbps
    \end{itemize}
    \item \textit{Tasks}:
    \begin{itemize}
        \item Task1: 8 CPUs, 32 GB RAM, [GPU], 3h, 10GB, Dependencies: []
        \item Task2: 4 CPUs, 16 GB RAM, [CPU], 2h, 5GB, Dependencies: [Task1]
        \item Task3: 16 CPUs, 64 GB RAM, [CPU, SSD], 5h, 20GB, []
        \item Task4: 8 CPUs, 32 GB RAM, [CPU], 4h, 15GB, Dependencies: [Task2, Task3]
    \end{itemize}
    \item \textit{Objectives}: Minimize makespan, balance load, efficient use of multi-feature nodes.
    \item \textit{Constraints}: No over-allocation, respect dependencies, account for data transfer if tasks assigned to different nodes.
\end{itemize}

We use data transfer time on purpose to increase the use case to a realistic scenario - when the ratio of output data increases, then it becomes an important factor. We can see if the model grasp such nuances for the experimental setup.

\begin{figure}[h!]
\centering
\begin{tikzpicture}[node distance=0.5cm and 1cm, every node/.style={draw, fill=blue!7, rounded corners, minimum width=1.6cm, minimum height=0.8cm, font=\normalsize}, >=Stealth]

% Nodes
\node (T1) {Task1};
\node[below=of T1] (T2) {Task2};
\node[right=of T1] (T3) {Task3};
\node[right=of T2] (T4) {Task4};

% Edges
\draw[->, thick] (T1) -- (T2);
\draw[->, thick] (T2) -- (T4);
\draw[->, thick] (T3) -- (T4);

\end{tikzpicture}
\caption{DAG for the sample HPC workflow}%: Task1 $\rightarrow$ Task2 $\rightarrow$ Task4; Task3 $\rightarrow$ Task4.}
\label{fig:sample_dag}
\end{figure}
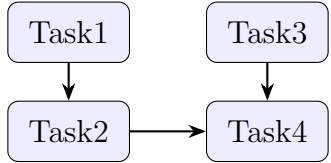

\subsection{Evaluation Metrics}
Following metrics are considered for the evaluation of the models 1) \textit{Makespan}: Total completion time of all tasks, 2) \textit{Constraint Violations}: Resource, feature, and dependency adherence, 3) \textit{Node Utilization Balance}: Degree of load balancing, and 4) \textit{Solution Explainability}: Clarity and logic of reasoning (qualitative).

The baseline makespan and mapping used for comparison correspond to the analytically verified optimum of 9\,h\,20\,s, computed manually. This ensures that evaluation reflects each model’s reasoning fidelity rather than differences in algorithmic implementation.

\subsection{Experimental Procedure}

Each scenario is evaluated manually for a baseline solution then by the solvers like a MILP optimizer (PuLP-based, extended for data transfer), an HEFT heuristic baseline and LLMs via API, with chain-of-thought prompting. For this sample scenario, we get minimum makespan of 9 hours and 20 seconds with 100\% task throughput (all four tasks assigned) and with resource utilization of two out of three nodes, as we prioritized makespan over optimum resource usage.

%Imagine a newcomer uses the simple prompt...
%Julian's Prompt...
For LLMs, at first we used a simple prompt, but during testing various prompts, we were assured to see already good reflection of the models and sometimes a correct results. This shows the results were often incomplete due to the ambiguity of the instructions, so we further refined the prompt as shown below:%(Section~\ref{sssec:LLM_Prompt})

%\begin{quote}
\subsubsection*{Prompt given to each LLM for benchmarking}
\label{sssec:LLM_Prompt}

%\begin{minipage}{\textwidth}
%\scriptsize
\begin{lstlisting}
You are an expert high-performance computing (HPC) workload scheduler.

Below is a description of a heterogeneous HPC system and a set of computational tasks (a workflow) to be scheduled. Your objectives are:
- Each task should be assigned to a node, specifying the start and end time for each task.
- Ensure that all node resource limits (CPUs, RAM), task feature requirements, dependencies, and system specifications are satisfied.
- Consider both the computational resources and the speed of data transfer between nodes when assigning tasks, as data movement and transfer delays directly impact total makespan.
- Maximize parallel execution of independent tasks whenever node resources allow.
- Important: If a task depends on one or more other tasks that ran on different nodes, the output data transfer for each dependency may begin only after the producing (parent) task has fully completed. The dependent task may not start until all required data from its dependencies has arrived at its assigned node.
- Data transfer delay must always be added if a task's dependency ran on a different node. Use following formula to calculate: Data Transfer Time (in hours) = ((Data Size (GB) * 8 bits/Byte) / (Min(Source Node Data Rate (Gbps), Destination Node Data Rate (Gbps)))) / 3600 seconds/hour)
- For example: if a 20GB output must be sent to a node with 10 Gbps, the transfer takes 16 seconds.
- Multi-feature nodes (e.g., GPU, SSD) should be reserved for tasks that need those features; avoid assigning tasks to such nodes if not necessary.
- Provide step-by-step reasoning for each assignment, including explicit explanations for any trade-offs or data transfer considerations.
- (Optional) Generate code (Python or pseudocode) that would automate or verify the final schedule.

SYSTEM NODES
- NodeA: 32 CPUs, 128 GB RAM, Features: [CPU, GPU], Data Transfer Rate: 10 Gbps
- NodeB: 64 CPUs, 256 GB RAM, Features: [CPU], Data Transfer Rate: 5 Gbps
- NodeC: 16 CPUs, 64 GB RAM, Features: [CPU, SSD], Data Transfer Rate: 2 Gbps

WORKLOAD TASKS
- Task1: Needs 8 CPUs, 32 GB RAM, Features: [GPU], Duration: 3h, Data Output: 10GB, Dependencies: []
- Task2: Needs 4 CPUs, 16 GB RAM, Features: [CPU], Duration: 2h, Data Output: 5GB, Dependencies: [Task1]
- Task3: Needs 16 CPUs, 64 GB RAM, Features: [CPU, SSD], Duration: 5h, Data Output: 20GB, Dependencies: []
- Task4: Needs 8 CPUs, 32 GB RAM, Features: [CPU], Duration: 4h, Data Output: 15GB, Dependencies: [Task2, Task3]

OPTIMIZATION OBJECTIVES: 
1. Minimize makespan, 
2. Optimum Resource Utilization.

CONSTRAINTS: 
1. One task should be assigned to only one node.
2. Task should be assigned within the node capacity.
3. Task feature request should be respected.
4. Task dependency should be respected.
5. Data transfer time should be considered in case if dependent tasks are assigned to different nodes.

EXPECTED OUTPUT FORMAT
For each task, output a table with:
- Task ID
- Assigned Node
- Start Time (considering dependencies and any transfer)
- End Time
- If data transfer was needed (yes/no, specify time and data amount)
- Short explanation

After the table, provide:
- Overall schedule makespan
- Any generated code (if produced)
- Any issues or assumptions you made
- (Optional) If possible, also provide an alternative schedule with a less efficient mapping and its resulting makespan for comparison.
\end{lstlisting}

\subsubsection{Explanation of Baseline Map-Schedule and Makespan:}

%The scheduling graph illustrate the optimal solution for our benchmark scenario. 
Table~\ref{tab:all_combos_makespan} enumerates all feasible task--node combinations for $T_2$ and $T_4$, 
since $T_1$ (GPU) and $T_3$ (SSD) have fixed node assignments. 
The makespan is determined by the longest dependency chain and any required inter-node data transfers. 
As shown, mapping $T_2$ to NodeA and $T_4$ to NodeC yields the minimum overall completion time of 
\textbf{9\,h\,20\,s}, as it avoids the large 20\,GB transfer from NodeC to NodeA. 
This configuration forms the analytical optimum and is used as the ground truth for evaluating LLM results.

\begin{table}[h!]
\centering
\caption{All task--node combinations for $T_2$ and $T_4$ with transfer delays and resulting makespan.
$T_1\to A$ (GPU) and $T_3\to C$ (SSD) are fixed. Transfer times use $(\text{GB}\cdot 8)/\min(\text{Gbps}_\text{src},\text{Gbps}_\text{dst})$.}
\label{tab:all_combos_makespan}
\scriptsize
\renewcommand{\arraystretch}{1.15}
\resizebox{\textwidth}{!}{
\begin{tabular}{c c | r r r | c c}
\hline
\textbf{$T_2$ node} & \textbf{$T_4$ node} 
& \textbf{$T_1\!\to\!T_2$ (s)} & \textbf{$T_2\!\to\!T_4$ (s)} & \textbf{$T_3\!\to\!T_4$ (s)} 
& \textbf{$T_4$ start (h:m:s)} & \textbf{Makespan (h:m:s)} \\ \hline

A & A & 0  & 0  & 80 & 5:01:20 & 9:01:20 \\
A & B & 0  & 8  & 80 & 5:01:20 & 9:01:20 \\

\rowcolor{gray!12}
\textbf{A} & \textbf{C} & \textbf{0} & \textbf{20} & \textbf{0} & \textbf{5:00:20} & \textbf{9:00:20} \\

\hline
B & A & 16 & 8  & 80 & 5:01:20 & 9:01:20 \\
B & B & 16 & 0  & 80 & 5:01:20 & 9:01:20 \\
B & C & 16 & 20 & 0  & 5:00:36 & 9:00:36 \\
\hline
C & A & 40 & 20 & 80 & 5:01:20 & 9:01:20 \\
C & B & 40 & 20 & 80 & 5:01:20 & 9:01:20 \\
C & C & 40 & 0  & 0  & 5:00:40 & 9:00:40 \\ \hline
\end{tabular}
}
\end{table}

\vspace{-0.5em} {\footnotesize
Notes: $T_2$ starts at 3\,h plus $T_1\!\to\!T_2$ transfer if off-node (10\,GB: A$\to$B 16\,s, A$\to$C 40\,s). $T_4$ begins once both $T_2$ and $T_3$ outputs arrive at its assigned node. The optimal mapping is \textbf{$T_2\to A$, $T_4\to C$}, achieving a minimal makespan of \textbf{9\,h\,20\,s} by keeping $T_3$ local on NodeC and transferring only 5\,GB from NodeA (20\,s delay).}

\section{Findings and Discussion}
\label{sec:findinganddiscussion}

\subsection{Findings}

In evaluating LLMs alongside traditional optimization approaches~\cite{sharma2025workflow}, several consistent patterns emerged. 
Table~\ref{tab:assessment} summarizes their behavior across all 21 publicly available models tested under a 30\,s response-time threshold. 
Every model successfully produced a feasible schedule, achieving full task coverage and balanced node utilization. 
However, the resulting schedules often varied across repeated runs with identical prompts, reflecting the intrinsic stochasticity of generative inference. 
This variability indicates that current LLM-generated schedules should be interpreted as probabilistic, near-feasible solutions rather than deterministic optima.

Table~\ref{tab:assessment} summarizes the comparative performance of state-of-the-art large language mod- els (LLMs) and code-oriented AI systems on the canonical HPC scheduling scenario. We assess the results based on what it produced denoted by $+$ (correct), $0$ (not given), $-$ (wrong), NA (not able).

\begin{table*}[h!]
\centering
\caption{Qualitative assessment of LLM performance on HPC workload mapping and scheduling benchmarks (inference parameters: temperature\,=\,0.5, top\_p\,=\,50\%). 
Columns denote: \textit{Makespan} (total workflow completion time), \textit{Task Throughput} (percentage of tasks successfully scheduled), \textit{Node Utilization} (efficient resource usage), \textit{Constraint Adherence} (compliance with resource, dependency, and transfer limits), \textit{Reasoning} (logical step-by-step justification), \textit{Explanation} (clarity of narrative output), \textit{Code} (presence and functionality of generated validation code), and \textit{Solution Explainability} (overall transparency and interpretability of the result).}
\label{tab:assessment}
\resizebox{\textwidth}{!}{%
\begin{tabular}{l|c||cc|c|c|c|c|c}
    \hline
    \textbf{Model} & \textbf{Optimum} 
    & \textbf{Task} & \textbf{} 
    & \textbf{Constraint} 
    & \textbf{Reasoning} 
    & \textbf{Explanation} 
    & \textbf{Working}  
    & \textbf{Response} \\

    & \textbf{Makespan} & \textbf{Throughput} & \textbf{Node Util.} 
    & \textbf{Adherence}
    & & \textbf{} & %\textbf{}
    \textbf{Code} & 
    \textbf{time} \\
    \hline

Llama 3.1 8B                        & 9h 32s      & + & + & - & - & + & - & + \\
Gemma 3 27B                         & 9h 1m 28s   & + & + & - & - & + & + & + \\
InternVL2.5 8B MPO                  & 20h 16s     & + & + & - & - & + & + & + \\

Qwen 3 32B                          & 9h 1m 20s      & + & + & + & + & - & + & - \\

DeepSeek R1                         & 11h         & + & + & - & - & + & - & - \\

Llama 3.1 Sauerkraut70B             & 9h 16m 32s  & + & + & - & + & + & + & + \\
Mistral Large                       & 9h 1m 28s   & + & + & - & + & + & + & - \\
Codestral 22B                       & 12h 32m     & + & + & - & - & + & + & - \\
%E5 Mistral 7B                      &             &   &   &   &   &   &   &   \\
Qwen 2.5 VL 72B                     & 9h          & + & + & - & - & + & + & - \\
Qwen 2.5 Coder 32B                  & 9h 4s       & + & + & - & - & + & + & + \\
o3                                  & 9h 60s      & + & + & + & + & + & + & - \\
o3-mini                             & 9h 20s      & + & + & + & + & - & - & + \\
GPT-4o                              & 9h 32s      & + & + & - & + & + & + & + \\
GPT-4o Mini                         & 9h 8s       & + & + & - & - & + & + & + \\
GPT-4.1                             & 9h 1m 20s   & + & + & + & + & + & + & + \\
GPT-4.1 Mini                        & 9h 20s      & + & + & + & + & + & + & + \\
\hline
Gemini Flash 2.5                    & 9h 1m 20s   & + & + & + & + & - & + & - \\
Gemini Pro 2.5                      & 9h 20s      & + & + & + & + & + & + & + \\
Microsoft Copilot                   & 9h 1m 20s   & + & + & + & + & + & + & + \\
Grok                                & 9h 1m 20s   & + & + & + & + & + & + & + \\
Claue Sonnet 4                      & 9h 1m 20s   & + & + & + & + & + & + & + \\
\hline 
\end{tabular}
}

\end{table*}

Among the 21 models evaluated, three (\texttt{o3-mini}, \texttt{GPT-4.1 Mini}, and \texttt{Gemini Pro 2.5}) precisely reproduced the analytically verified optimum makespan of 9\,h\,20\,s while fully satisfying all resource, feature, and dependency constraints. 
Twelve additional models, including \texttt{o3}, \texttt{GPT-4.1}, \texttt{Qwen 3 32B}, \texttt{Qwen 2.5 VL 72B}, \texttt{Qwen 2.5 Coder 32B}, \texttt{Mistral Large}, \texttt{Gemma 3 27B}, \texttt{GPT-4o Mini}, \texttt{Gemini Flash 2.5}, \texttt{Microsoft Copilot}, \texttt{Grok}, and \texttt{Claude Sonnet 4}, achieved near-optimal makespans within one to two minutes of the reference. 
Models such as \texttt{Llama 3.1 8B}, \texttt{Llama 3.1 Sauerkraut 70B}, \texttt{DeepSeek R1}, \texttt{Codestral 22B}, and \texttt{InternVL2.5 8B MPO} produced suboptimal results or violated one or more constraints, primarily due to miscalculations in data-transfer delays or serialized execution of independent tasks. 
Across all models, task throughput and node utilization reached 100\%, but only about half maintained strict constraint compliance. 
19 models generated partially executable validation code, and 18 provided coherent step-by-step reasoning, demonstrating strong interpretability even when minor arithmetic or dependency errors occurred.

A key differentiator among models was \textit{constraint adherence and makespan accuracy}. 
While most produced feasible mappings, only a few, particularly \texttt{o3}, \texttt{o3-mini}, \texttt{GPT-4.1 Mini}, and \texttt{Gemini Pro 2.5}, consistently respected dependency order and transfer timing, yielding the true optimal makespan (9\,h\,20\,s). 
Others, including \texttt{GPT-4.1}, \texttt{Qwen 3 32B}, and \texttt{Mistral Large}, were near-optimal but occasionally underestimated transfer delays or overlooked subtle resource conflicts. 
These deviations highlight a persistent limitation in current LLMs’ ability to precisely enforce hard temporal and resource constraints.

Regarding \textit{mapping efficiency and node utilization}, higher-performing models effectively leveraged resource locality and hardware features, co-locating dependent tasks and assigning workloads to appropriate accelerators. 
In contrast, weaker models often ignored opportunities for parallelism or failed to minimize data transfer overhead, directly increasing the makespan.

A notable common strength across nearly all advanced models was their \textit{reasoning transparency and code generation}. 
Most provided interpretable, step-by-step rationales and readable verification code. 
Models such as \texttt{o3}, \texttt{GPT-4o}, and \texttt{Mistral Large} demonstrated clear, auditable reasoning traces even when minor logical inconsistencies occurred. 
Interestingly, no consistent correlation was observed between model size or response latency and scheduling accuracy, larger, slower models often performed comparably to smaller, faster ones, suggesting that prompt comprehension and reasoning structure, rather than scale, govern LLM performance in this task.

\begin{figure}[h!]
  \centering
  \begin{tikzpicture}
    \begin{axis}[
      ybar,
      bar width=12pt,
      enlarge x limits=0.15,
      ylabel={\# Models Succeeding},
      symbolic x coords={
        Throughput, NodeUtil, Constraint, Reasoning,
        Explanation, Code, Response
      },
      xtick=data,
      x tick label style={rotate=45, anchor=east},
      ymin=0, ymax=25,
      nodes near coords,
      nodes near coords align={vertical},
      ]
      \addplot+[fill=blue!60] coordinates {
        (Throughput,21)
        (NodeUtil,21)
        (Constraint,10)
        (Reasoning,13)
        (Explanation,18)
        (Code,19)
        (Response,14)
      };
    \end{axis}
  \end{tikzpicture}
  \caption{Number of models (out of 21) that succeeded on each qualitative metric.}
  \label{fig:success_counts}
\end{figure}
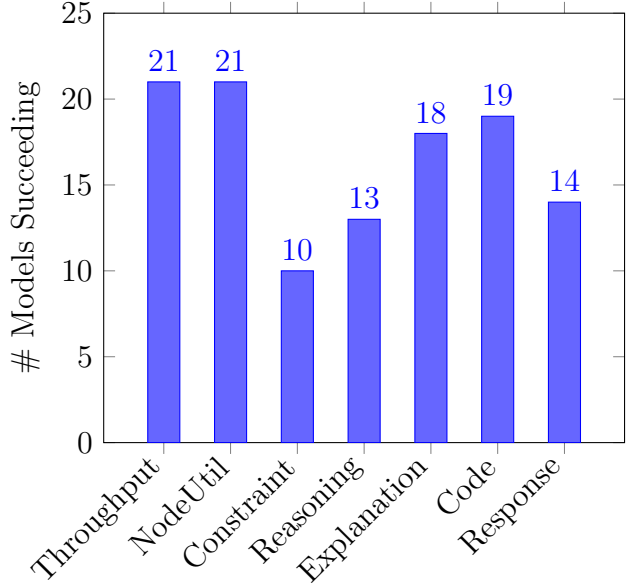

Figure~\ref{fig:success_counts} summarizes the number of models (out of 21) that succeeded on each qualitative metric. 
All models achieved perfect scores in \emph{Task Throughput} and \emph{Node Utilization}, confirming their consistent ability to complete all tasks and allocate resources efficiently. 
Only ten models demonstrated full \emph{Constraint Adherence}, indicating that nearly half occasionally violated dependency or resource limits. 
Performance on the \emph{Reasoning} metric was moderate (thirteen models), reflecting variation in the depth and accuracy of step-by-step logic. 
In contrast, high success rates were observed for \emph{Explanation} (eighteen models) and \emph{Code Generation} (nineteen models), showing strong capability in producing clear rationales and functional validation scripts. 
\emph{Response Time} success was achieved by fourteen models, suggesting that while most responded within acceptable latency, a notable fraction remains slower or less efficient in inference. 
Overall, the distribution indicates that although LLMs excel in generating feasible mappings and interpretable reasoning artifacts, their constraint-enforcement accuracy and logical consistency still require improvement.

%Qwen 3 32B - stopped after some unfinished reasoning.
%Great to understand, considered network transfer time.

%Findigs:
%Some model stops during assessment, we then had to say "continue".

\subsection{Illustrative Reasoning Examples}

To qualitatively demonstrate model interpretability and reasoning diversity, this section presents two representative examples of how LLMs reasoned about mapping and scheduling constraints from natural-language input. 
While most models produced feasible schedules, their internal reasoning strategies varied significantly, some structured the problem symbolically, others expressed it procedurally in text.

\subsubsection{Llama 3.1 8B ,  Symbolic Graph Representation}
\texttt{Llama~3.1~8B} modeled the HPC system as a directed graph using \texttt{networkx}, assigning resource attributes to nodes and defining task dependencies as edges. 
This demonstrates an emerging capability to abstract the scheduling problem into a formal structure without explicit instructions or templates.

\begin{lstlisting}[
    basicstyle=\small,breaklines=true,
    caption={Excerpt from Llama 3.1 8B: Graph-based abstraction of system resources and tasks.}
]
G = nx.DiGraph()
G.add_node("NodeA", cpus=32, ram=128, features=["GPU"])
G.add_edge("NodeA","NodeA", task="Task1", features=["GPU"], duration=3)
\end{lstlisting}

This behavior suggests that some models attempt to reconstruct implicit optimization structures, transforming textual inputs into symbolic, machine-readable representations.

\subsubsection{Qwen 3 32B ,  Natural-Language Reasoning}
In contrast, \texttt{Qwen~3~32B} provided explicit natural-language reasoning to explain each decision step. 
It tracked remaining resources after every assignment and justified task placements based on feature requirements and dependencies:

\begin{quote}
\textit{First, assign Task1 to NodeA. That uses 8 CPUs and 32 GB RAM, leaving 24 CPUs and 96 GB free. 
Task3 requires SSD, so assign it to NodeC, which fully utilizes its 16 CPUs and 64 GB RAM.}
\end{quote}

This linguistic reasoning trace highlights an interpretable, procedural approach, making the decision process auditable by humans.

\noindent
Together, these examples reveal two complementary reasoning modalities, symbolic graph construction and narrative explanation, illustrating how current LLMs conceptualize scheduling constraints. 
Such interpretability reinforces their potential as explainable co-pilots in hybrid optimization systems rather than opaque black-box solvers.

\subsection{Discussion}

This study provides a clear snapshot of current LLM reasoning capability in HPC workload mapping and scheduling in heterogeneous landscape. 
All 21 models generated feasible task–node mappings and coherent explanations, but their probabilistic reasoning led to output variability, identical prompts sometimes produced different schedules or makespans. 
Unlike deterministic optimizers, LLMs approximate constraint reasoning rather than enforcing it, making their results interpretable yet non-guaranteed.

Most models achieved perfect task throughput and balanced node utilization, but only \textbf{10 of 21} satisfied all constraints without violation. 
Frequent errors included incorrect transfer-time arithmetic, ignored dependency ordering, or minor resource over-allocations. 
Only \textbf{3 models} (\texttt{o3}, \texttt{o3-mini}, \texttt{GPT-4.1 Mini}) exactly reproduced the analytical optimum of 9 h 20 s, while \textbf{12 others} were near-optimal (within one to two minutes). 
These deviations show that LLMs reason plausibly about parallelism and critical paths but lack consistent enforcement of temporal and resource constraints.

Scalability remains limited: as prompts grow more complex, some models truncate reasoning or omit constraints, yielding incomplete schedules. 
Still, their transparency stands out, LLMs articulate decision logic, trade-offs, and assumptions in natural language, an ability absent from mathematical solvers. 
This interpretability positions them as effective \textit{co-pilots} that can draft candidate mappings, explain heuristic choices, and translate human objectives into structured optimization inputs.

\subsubsection*{Research Question Analysis}

\noindent\textbf{RQ1: Understanding constraints:}  
LLMs generally comprehend resource and feature requirements but partially fail on inter-task dependencies and data-transfer rules. 
They rely on explicit prompt phrasing to maintain constraint consistency.

\noindent\textbf{RQ2: Optimality and makespan accuracy:}  
A minority matched the analytical optimum, and most remained near-optimal. 
Variability stems from approximate arithmetic and non-deterministic reasoning, not from lack of understanding.

\noindent\textbf{RQ3: Failure patterns and implications:}  
Typical issues include unit conversion errors, premature task starts, conservative serialization, and context loss in long prompts. 
These patterns suggest using LLMs as explainable front-ends, complemented by symbolic solvers or rule-based validators, to achieve reliable hybrid scheduling.

\noindent
Overall, LLMs already approximate optimal solutions with strong interpretability for small, well-defined workloads. 
Their role is best envisioned not as autonomous optimizers but as reasoning assistants embedded within human-in-the-loop or hybrid optimization pipelines.

\section{Conclusion and Future Work}
\label{sec:conclusionandfuturework}

This study systematically evaluated the intrinsic reasoning capabilities of 21 Large Language Models (LLMs) for workload mapping and scheduling in heterogeneous HPC systems using only natural-language descriptions. 
A manually derived analytical optimum of 9\,h\,20\,s served as the ground-truth reference for correctness and constraint validation. 
Out of the 21 evaluated models, three (\texttt{o3-mini}, \texttt{GPT-4.1 Mini}, and \texttt{Gemini Pro 2.5}) exactly reproduced the analytical optimum, twelve achieved near-optimal makespans within two minutes, and six produced suboptimal or inconsistent schedules. 
Overall, approximately three-quarters of the models demonstrated the ability to reconstruct near-optimal reasoning paths for this small-scale HPC scheduling task.

\noindent
These results confirm that LLMs can comprehend and reason over complex resource, feature, and dependency relationships described purely in natural language, addressing \textbf{RQ1}. 
In relation to \textbf{RQ2}, most models derived feasible and interpretable schedules, though only a small subset achieved exact optimality due to probabilistic reasoning and arithmetic inaccuracies. 
Regarding \textbf{RQ3}, observed failure patterns included unit-conversion errors, overlooked transfer delays, and premature dependency resolution, emphasizing the need for explicit constraint reinforcement and validation mechanisms.

Despite these limitations, the findings establish that LLMs can function as \textit{explainable reasoning co-pilots} capable of generating feasible schedules, human-readable justifications, and verification code. 
Their interpretability and flexibility make them promising components for hybrid scheduling workflows that combine language-based reasoning with deterministic optimization engines.

\noindent
Future work will expand this benchmark toward larger and more complex workflow graphs, integrate automated constraint-verification pipelines, and explore hybrid LLM architectures. 
In such systems, LLMs could provide natural-language reasoning and constraint translation, while formal solvers ensure quantitative precision. 
This hybrid paradigm represents a key step toward scalable, explainable, and reliable AI-assisted workload scheduling in heterogeneous HPC environments.

\section*{Acknowledgments}
This work was supported by the University of Göttingen. The authors gratefully acknowledge the Gesellschaft für wissenschaftliche Datenverarbeitung mbH Göttingen (GWDG) for providing computational resources and technical support. We also thank our colleagues and peers in the HPC and AI research community for their constructive feedback and valuable discussions. This research was funded by the BMBF KISSKI Project under grant number 01 IS 22 093 A-E.

\bibliographystyle{plain}

\end{document}